\documentclass[conference]{IEEEtran}
\ifCLASSINFOpdf
\else
\fi
\usepackage{times}
\usepackage{graphicx}
\usepackage{amsmath}

\newcommand{\remove}[1]{}

\newcommand{\figref}[1]{Figure~\ref{#1}}

\title{Leveraging User Diversity to Harvest Knowledge on the Social Web}
\begin{document}
%

\author{
\IEEEauthorblockN{Jeon-Hyung Kang}
\IEEEauthorblockA{Information Sciences Institute\\
University of Southern California\\
Marina del Rey, California 90292\\
Email:jeonhyuk@usc.edu}
\and
\IEEEauthorblockN{Kristina Lerman}
\IEEEauthorblockA{Information Sciences Institute\\
University of Southern California\\
Marina del Rey, California 90292\\
Email: lerman@isi.edu}
}


%


\maketitle

\begin{abstract}
Social web users are a very diverse group with varying interests, levels of expertise, enthusiasm, and expressiveness. As a result, the quality of content and annotations they create to organize content is also highly variable.
While several approaches have been proposed to mine social annotations, for example, to learn folksonomies that reflect how people relate narrower concepts to broader ones, these methods treat all users and the annotations they create uniformly. We propose a framework to automatically identify experts, i.e., knowledgeable users who create high quality annotations, and use their knowledge to guide folksonomy learning. We evaluate the approach on a large body of social annotations extracted from the photosharing site Flickr. We show that using expert knowledge leads to more detailed and accurate folksonomies. Moreover, we show that including annotations from non-expert, or novice, users leads to more comprehensive folksonomies than experts' knowledge alone.
\end{abstract}


%
\IEEEpeerreviewmaketitle

\section{Introduction}
Knowledge production is no longer solely in the hands of professionals: on many Social Web sites ordinary people create and annotate a wide variety of content. On the social photosharing site Flickr ({http://flickr.com}), for example, users can publish photographs, tag them with descriptive keywords, such as \texttt{insect} or \texttt {macro},  and organize them within personal directories. While an individual's annotations express her particular world view, collectively social annotations provide valuable evidence for harvesting social knowledge, including $folksonomies$ ($folk+taxonomies$) that show how people relate broader concepts to narrower ones. Social knowledge is idiosyncratic and may at times conflict with knowledge expressed in professionally curated taxonomies. For example, many people consider spiders to be \texttt{insects}, at odds with the Linnean taxonomy of living organisms. However, such knowledge is necessary to make sense of and leverage user-generated content on the Social Web. Thus, to find all images of \texttt{spiders}, you will sometimes have to look for \texttt{insects}.

Recently, Plangprasopchok \emph{et al.}~\cite{Plangprasopchok11wsdm} proposed a method to learn {folksonomies} by integrating structured annotations from many users, specifically, personal directories created by individual Flickr users to organize their photos. The method extends affinity propagation~\cite{apscience07} to use structural information to concurrently combine many shallow personal directories into a larger common taxonomy. The method assumes that the \emph{quality} of annotation from all users is the same. However, Social Web users are highly diverse and vary in their degree of expertise and expressiveness. Knowledgeable users create high quality, detailed annotations,  often using technical terms. They specify intermediate concepts within multi-level directories, e.g., linking \texttt{jumping spider} to \texttt{spiders} to \texttt{arachnids} to \texttt{invertebrates}. We call such users \emph{experts}. Novice users, on the other hand, are far less expressive, creating shallow directories that jump granularity levels, e.g.,  linking \texttt{spiders} to \texttt{bugs}. Using experts' knowledge enables us to learn more accurate and detailed folksonomies. 

Diversity is important for groups and organizations~\cite{Stirling}. It can lead to better group decision making and organizational robustness~\cite{Surowiecki2005}, as long as individual knowledge and opinions are aggregated correctly~\cite{Page:2007:DPD:1296125}. Hence, identifying experts from the content they create, or from recommendations of other people, has been an active research area. Previous works used natural language analysis~\cite{Maybury02,balog2006formal} and topic modeling~\cite{deng2009formal} techniques to identify experts from the text of documents they created, often combining it with analysis of the structure of links within an organization~\cite{zhang2010expert,Davitz}. Annotations on the Social Web can help identify diverse classes of users. However, while previous researchers classified users based on their annotation practices~\cite{koerner2010stop}, they did not attempt to automatically distinguish expert from novice users.

In this paper we propose methods to automatically identify expert users who provide high quality annotations and leverage their knowledge in folksonomy learning.
First, in Section~\ref{sec:specialist}, we describe and evaluate a method that examines structured annotations to automatically identify expert users. Specifically, our method analyzes the structure and content of personal directories created by Flickr users. In Section~\ref{sec:rap} we extend the inference method of Plangprasopchok et al.~\cite{Plangprasopchok11wsdm} to use experts' knowledge to guide the folksonomy learning process. In Section~\ref{sec:validation} we show that the inference method that exploits user diversity by putting greater weight on annotations created by experts can learn more accurate and detailed folksonomies than one that ignores diversity. Surprisingly, however, we show that while experts' knowledge is required to learn more accurate folksonomies, novice knowledge is needed to learn more complete folksonomies.  We also carry out a detailed investigation of the robustness of our method.


%

\section{Identifying Expert Users}
\label{sec:specialist}
Experts are knowledgeable individuals who can answer questions within organizations and generate high quality data. Identifying such people is an important research topic in data mining, management science, and social network analysis. Researchers have proposed a variety of algorithms for automatic expert identification, including language~\cite{balog2006formal}, probabilistic topic-based~\cite{deng2009formal} and statistical~\cite{Maybury02} models and network analysis tools~\cite{Campbell03expertiseidentification,Davitz}, that identify experts based on the documents or email messages they exchange within organizations. Hybrid approaches that combine topics and relationships between users~\cite{zhang2010expert} have also been explored.

Expert identification is even more important for mining user-generated content, since Social Web users form an extremely diverse group, with widely varying levels of expertise and enthusiasm for different topics. As a result, the quality of data they create also varies tremendously. One way to differentiate data quality is by identifying expert users. We extend the features used to measure diversity in groups~\cite{Stirling} and use them within a supervised expert classification method. The features measure users' expertise based on the \emph{structure of annotations} they create. Unlike previous works that examined (textual) data people create, our method looks directly at knowledge structures they express through annotations.

\subsection{Structured Annotations}
Social web sites allow users to annotate content they create or share with others. In addition to tagging content, some sites also allow users to organize it hierarchically. {Del.icio.us} \remove{(\texttt{http://del.icio.us/})} users can group related tags into bundles, and {Flickr} users can group related photos into \emph{sets} and then group related sets in \emph{collections}, thereby creating personal directories to organize photos. The sites themselves do not impose any rules on the vocabulary or semantics of directories; in practice users employ them to represent \emph{relations between broader and narrower categories or concepts}.
Personal directories offer rich evidence for harvesting social knowledge and have been used to learn communal taxonomies of concepts, otherwise known as \emph{folksonomies}~\cite{Plangprasopchok10kdd,Plangprasopchok11wsdm}.

\remove{
\begin{figure*}[tbh]
\begin{tabular}{@{}c@{}c@{}}
 \includegraphics[height=1.3in]{maximillipede} &
\includegraphics[height=1.3in]{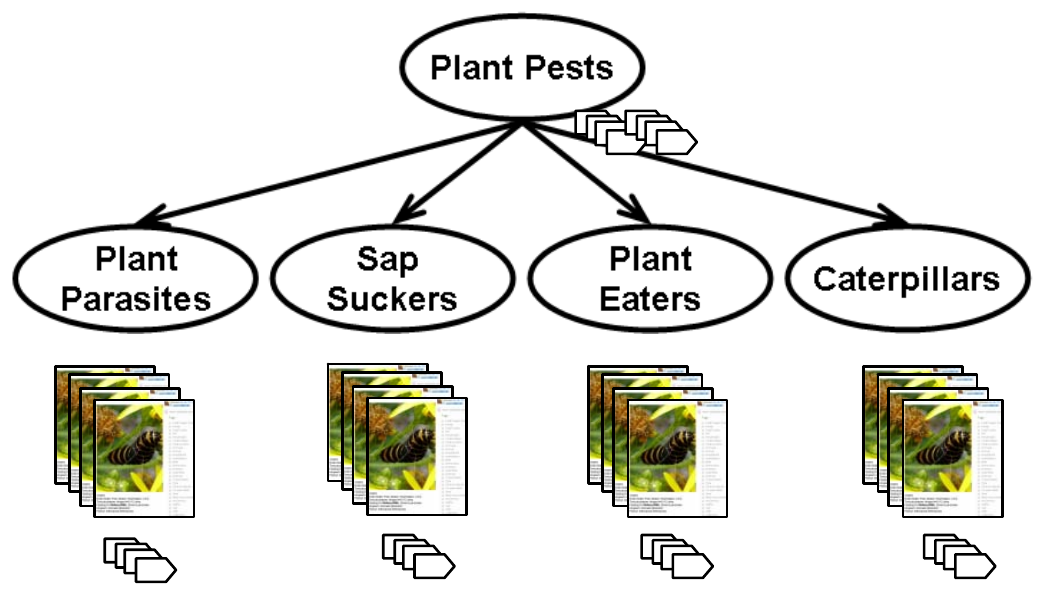}  \\
(a) & (b)
\end{tabular}
  \caption{Personal hierarchies specified by a Flickr user. (a) A collection \coll{Plant Pests} created by the user with its constituent sets, and  tags associated with an image in the \set{Caterpillars} set. (b) Schematic representation of the \coll{Plant Pests} hierarchy.}\label{fig:collections}
\end{figure*}
Figure~\ref{fig:collections} illustrates how this feature is implemented on the social photo-sharing site {Flickr}.
Figure~\ref{fig:collections}(a) shows one of the collections created by an avid naturalist on Flickr. This and other collections reflect the subjects she likes to photograph: \coll{Birds}, \coll{Mammals}, \coll{Plants}, \coll{Mushrooms \& Fungi}, \coll{Plant Diseases}, etc. The \coll{Plant Pests} collection contains sets \set{Plant Parasites}, \set{Sap Suckers}, \set{Plant Eaters}, and \set{Caterpillars}.
Each set contains one or more photos tagged by the user.
For example, a photograph in the set \set{Caterpillars} (\figref{fig:collections}(a)), is annotated with multiple tags, including (\term{Animal}, \term{Lepidoptera}, \term{Moth}, \term{larva}, \term{Caterpillar}), its color (\term{brown and orange}), context (\term{on porch}), and location (\term{King County}, \term{WA}, \term{North America}).
}

\begin{figure}[tb]
\begin{center}
\begin{tabular}{@{}c@{}}
\includegraphics[width=1.0\linewidth]{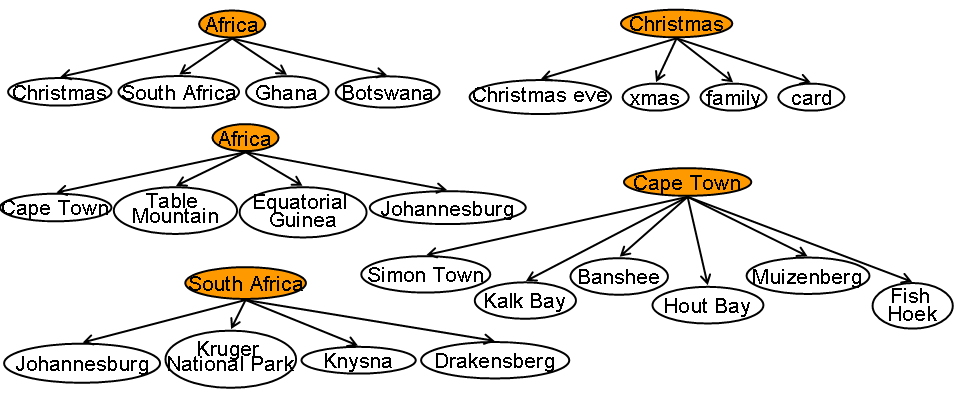}\\
(a) \\
\includegraphics[width=1.0\linewidth]{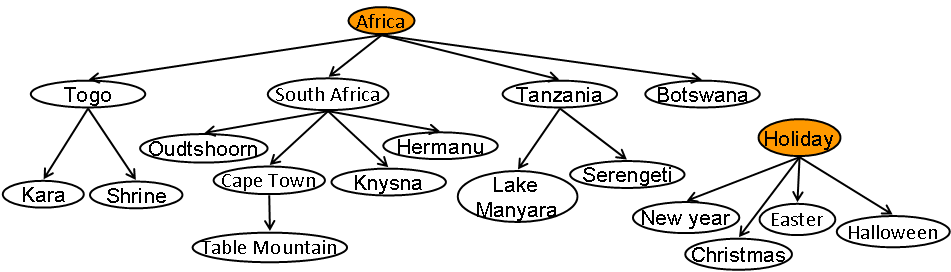}\\
(b)
\end{tabular}
\end{center}
\caption{Saplings created by (a) novice and (b) expert users.}
\label{fig:typeSPC}
\end{figure}

Following Plangprasopchok \emph{et al.}~\cite{Plangprasopchok10kdd} we call a directory a user creates to organize photos on Flickr a \emph{sapling}.\footnote{Saplings are not always tree-like. In these cases we convert them to trees.} The root node of the sapling corresponds to a user's collection, and inherits its name, while the leaves correspond to the collection's constituent sets (or other collections) and inherit their names. The photos the user assigns to a set are tagged, and we propagate these tags  to sets and to their parent collections.
While most users create shallow saplings consisting of a top-level collection and constituent sets (see Fig.~\ref{fig:typeSPC}(a)), others create detailed, multi-level hierarchies about a topic of interest (Fig.~\ref{fig:typeSPC}(b)). We call the latter users \emph{experts} and the former \emph{novice}.  By manually inspecting saplings created by Flickr users, we found that structure and semantic consistency are two  important factors distinguishing expert from novice users. Specifically, we have identified the following hallmarks of an expert:
\begin{itemize}
\item generally creates many saplings with distinct concepts
\item creates deep ($>2$ levels as in Fig.~\ref{fig:typeSPC}(b)) or broad saplings
\item provides top-level concepts that are meaningful to others. Overly-broad concepts, such as `life', `things', `misc', `all sets', etc., imply novice users
\item does not jump many levels, (e.g., attach `los angeles' to `world') nor mix concepts of different granularity level (e.g., `table mountain' and `equatorial guinea' are never siblings, as in Fig.~\ref{fig:typeSPC}(a)) 
\item does not create conflicts (e.g., attach `los angeles' to `journey' in one sapling while attaching `journey' to `los angeles' in another)
\item does not create multiple child concepts with same name (e.g., five `los angeles' sets under `journey').
\end{itemize}

\subsection{Features}
To automate expert identification, we convert the observations above into quantitative features. We divide the features into two classes: user-level and sapling-level features.


\subsubsection{User Features} Experts express a variety of concepts.

\textbf{User-Variety} measures the number of saplings ($N$) and \textbf{NumTwigs} the number of relations a user creates.

\textbf{User-Balance} measures how uniform the saplings are in size. We measure this by entropy
$B_U=(- \sum_{i}{p_i \ln p_i}) / \ln N,$
where $p_i$ is the number of nodes in sapling $i$ divided by the total number of nodes the user creates.

\textbf{User-Disparity} measures differences between concepts expressed in user's saplings~\cite{Stirling}. We compute disparity using Jensen-Shannon divergence between the tag distributions of the two saplings:
\begin{equation}
\label{eq:userdisparity}
\sum_{i,j} JS( \tau_{i} ||  \tau_{j})  = \sum_{i,j}  (0.5 D( \tau_{i} ||  \tau_{k}) + 0.5 D( \tau_{j} ||  \tau_{k})),
\end{equation}
\noindent where $\tau_{i}$ represents tag distribution of sapling i, $\tau_{k} =1/2 (\tau_{i} +\tau_{j})$ and $D(.)$ is Kullback-Leibler divergence.
\textbf{DisparityNormalized} simply divides the above measure by the number of nodes in the saplings.

\subsubsection{Sapling Features}
Experts express detailed knowledge in particulars topics, not necessarily all topics.

\textbf{Sapling-Variety} combines depth and breadth of the sapling: $V_S=\sum_{i=1:L}{ i \times n_i}$, where $L$ is the depth of the sapling and $n_i$ is the number of nodes at level $i$. This gives more credit to deeper representations if both saplings are equally large.

\textbf{Sapling-Balance} measures how balanced the sapling is at each level. We quantify balance by normalized entropy based on expected number of nodes at current level given the number of nodes at the previous level:
$B_S=1/L \times \sum_{i=1:L} \big( (- \sum_{j}{p^{i}_{j} \ln p^{i}_{j}}) / \ln (n_{i}) \big)$,
where $n_{i}$ is number of nodes at level $i$,  $p^{i}_{j}$ is proportion of children of $j$'th node at level $i$.  For example, if there are 4 nodes in level 1 with 3, 3, 1, 2 children respectively, then $n_{1}$ is 4, $p^{1}_{j}$  is (3/9, 3/9, 1/9, 2/9). To balance level 2, we need between two and three children per parent.

Several features measure concept consistency and node uniqueness. Inconsistency can be computed by the number of \textbf{conflicts} (i.e. attaching node $A$ to node $B$ in one sapling and $B$ to node $A$ in another sapling); \textbf{agreement} is quantified by how many users create the same parent-child relation; node (or twig) \textbf{uniqueness} is computed by the ratio of unique node names to the total number of nodes in the sapling.
Other features include sapling \textbf{depth}, \textbf{breadth}, number of \textbf{nodes} and terminal \textbf{leaves} it has, and the \textbf{ratio} of number of leaves to the total number of nodes in the sapling.

\textbf{Root-Diversity} is an important hallmark of experts.
Experts create generalizable knowledge using categories that are meaningful to others. A vague concept, such as `misc', `other', `things', will mean different things to different people. Consequently, there will be little agreement about the child concepts of such root nodes, with every user specifying a different child. There is far more agreement about the children of more specific concepts, such as `europe'.
We quantify the generalizability of a concept by the
the distribution of distinct child nodes across all users.

\begin{figure}[tb]
\begin{tabular}{@{}c@{}c@{}}
\includegraphics[width=0.5\linewidth]{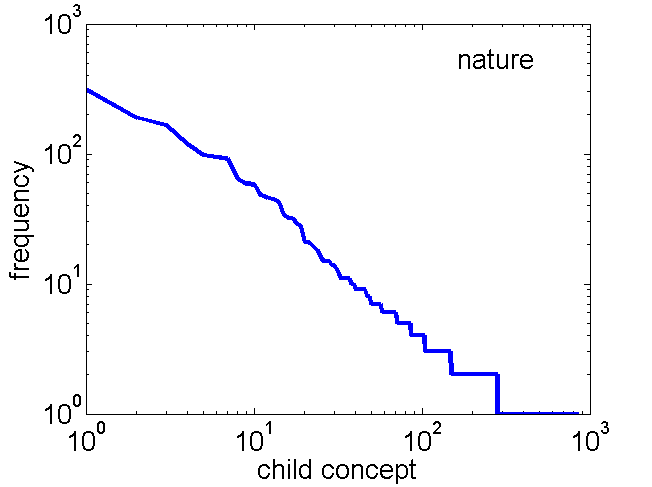} &
\includegraphics[width=0.5\linewidth]{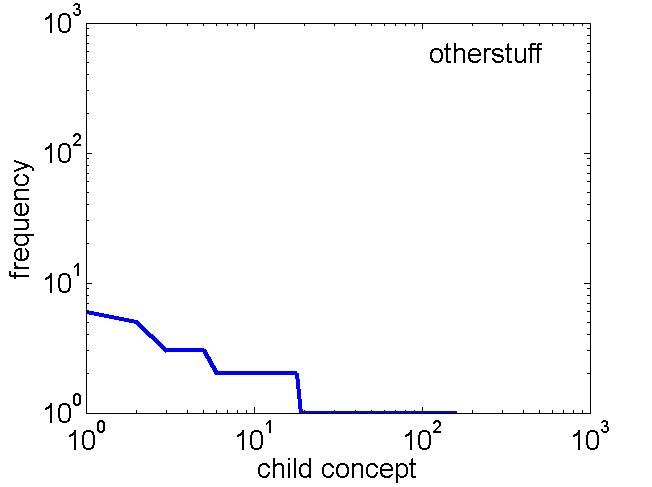}\\
(a) & (b)\\
\end{tabular}
\caption{Frequency distribution of distinct children of root nodes (a) `nature' and (b) `other stuff'.}
\label{fig:tword}
\end{figure}


Given a concept (sapling root), we extract all sub-concepts users have specified as children of this root. Figure~\ref{fig:tword} shows the distributions of unique children of the roots `nature' and `other stuff', sorted by frequency of occurrence. A peaked distribution (Fig.~\ref{fig:tword}(a)) indicates agreement among users about sub-concepts and implies that the root concept is meaningful to others. A flat distribution (Fig.~\ref{fig:tword}(b)) implies there is little agreement about the root concept, with practically each user expressing a different sub-concept. This indicates that the root concept is vague.  We quantify the peakedness of the distribution by measuring how many unique nodes are needed to cover 30\%, 50\% and 70\% of child nodes. For example, to cover 70\% of the distinct children of the root `europe', we need to look at 21.3\% of the most frequent children, while to cover the same fraction of children of `other stuff', we need to look at 64.6\% of the most frequent children.
Other root concepts in our data set that are meaningful to many users include `nature', `animal', `flower', `bird', `usa', `sport', while the vaguer, less meaningful concepts include `location', `subject', `everything else', `landscape', `random', `stuff', and `miscellaneous'.

Other features characterizing root diversity include the number of people who have created a root node with that name, and the number of unique children the root has over all users.

\subsection{Automatically Identifying Experts}
We collected saplings created by 7,121 Flickr users who were members of wildlife and nature photography public groups.
We {trained} a model to use the features above to automatically identify experts among these users. We trained the model on a small set of manually labeled data and used it to label a larger test set. 
We then examined and labeled new predictions made by the model, added them to the training set and retrained the model.
We iterated this self-training procedure on the unlabeled test data to discover new experts, and re-trained the model with the enriched data.

To create the initial training set, we asked three annotators to review saplings created by 200 Flickr users randomly selected from the set of 1000 who specified most relations. Annotators used the criteria above to identify experts. Each user's saplings were laid out hierarchically using yEd graph visualization tool. Annotators identified 20--45 experts among 200 users. We treated 19 experts all annotators agreed upon as positive, and the rest as negative, examples in the training set.

\begin{table}[tbh]
\centering
\setlength{\tabcolsep}{1pt}
\scriptsize{

\begin{tabular}{|c|c|c|c|c|c|c|c|c|c|c|c|}
\hline
\multicolumn{1}{|c|}{} & \multicolumn{9}{c|}{Training Set Cross Validation} & \multicolumn{1}{c|}{Training} & \multicolumn{1}{c|}{Positive}  \\
\cline{2-10}
\multicolumn{1}{|l|}{Iterations} & \multicolumn{3}{c|}{J48} & \multicolumn{3}{c|}{Random Forest} & \multicolumn{3}{c|}{LibSVM} & \multicolumn{1}{l|}{examples} & \multicolumn{1}{l|}{examples}\\
\cline{2-10}
& Pr & Re & F & Pr & Re & F & Pr & Re & F &  & \\
\hline
\multicolumn{1}{|c|}{1} & \multicolumn{1}{l|}{0.44}  & \multicolumn{1}{l|}{0.58}  & \multicolumn{1}{l|}{0.50}  & \multicolumn{1}{l|}{0.67}  & \multicolumn{1}{l|}{0.42}  & \multicolumn{1}{l|}{0.52}  & \multicolumn{1}{l|}{0.80} & \multicolumn{1}{l|}{0.63} & \multicolumn{1}{l|}{0.70} & \multicolumn{1}{c|}{200} & \multicolumn{1}{c|}{19} \\
\hline
\multicolumn{1}{|c|}{2} & \multicolumn{1}{l|}{0.56}  & \multicolumn{1}{l|}{0.50}  & \multicolumn{1}{l|}{0.53}  & \multicolumn{1}{l|}{0.61}  & \multicolumn{1}{l|}{0.50}  & \multicolumn{1}{l|}{0.55}  & \multicolumn{1}{l|}{0.76} & \multicolumn{1}{l|}{0.42} & \multicolumn{1}{l|}{0.54} & \multicolumn{1}{c|}{274} & \multicolumn{1}{c|}{38} \\
\hline
\multicolumn{1}{|c|}{3} & \multicolumn{1}{l|}{0.56}  & \multicolumn{1}{l|}{0.42}  & \multicolumn{1}{l|}{0.49}  & \multicolumn{1}{l|}{0.57}  & \multicolumn{1}{l|}{0.48}  & \multicolumn{1}{l|}{0.52}  & \multicolumn{1}{l|}{0.86} & \multicolumn{1}{l|}{0.57} & \multicolumn{1}{l|}{0.69} & \multicolumn{1}{c|}{292} & \multicolumn{1}{c|}{42} \\
\hline
\multicolumn{1}{|c|}{4} & \multicolumn{1}{l|}{0.51}  & \multicolumn{1}{l|}{0.55}  & \multicolumn{1}{l|}{0.53}  & \multicolumn{1}{l|}{0.50}  & \multicolumn{1}{l|}{0.53}  & \multicolumn{1}{l|}{0.45}  & \multicolumn{1}{l|}{0.88} & \multicolumn{1}{l|}{0.71} & \multicolumn{1}{l|}{0.78} & \multicolumn{1}{c|}{293} & \multicolumn{1}{c|}{42} \\
\hline
\multicolumn{1}{|c|}{5} & \multicolumn{1}{l|}{0.44}  & \multicolumn{1}{l|}{0.44}  & \multicolumn{1}{l|}{0.44}  & \multicolumn{1}{l|}{0.57}  & \multicolumn{1}{l|}{0.47}  & \multicolumn{1}{l|}{0.51}  & \multicolumn{1}{l|}{0.80} & \multicolumn{1}{l|}{0.58} & \multicolumn{1}{l|}{0.67} & \multicolumn{1}{c|}{297} & \multicolumn{1}{c|}{43} \\
\hline
\multicolumn{1}{|c|}{6} & \multicolumn{1}{l|}{0.50}  & \multicolumn{1}{l|}{0.41}  & \multicolumn{1}{l|}{0.45}  & \multicolumn{1}{l|}{0.54}  & \multicolumn{1}{l|}{0.34}  & \multicolumn{1}{l|}{0.42}  & \multicolumn{1}{l|}{0.84} & \multicolumn{1}{l|}{0.50} & \multicolumn{1}{l|}{0.63} & \multicolumn{1}{c|}{292} & \multicolumn{1}{c|}{43} \\
\hline
\multicolumn{1}{|c|}{7} & \multicolumn{1}{l|}{0.42}  & \multicolumn{1}{l|}{0.39}  & \multicolumn{1}{l|}{0.40}  & \multicolumn{1}{l|}{0.57}  & \multicolumn{1}{l|}{0.39}  & \multicolumn{1}{l|}{0.46}  & \multicolumn{1}{l|}{0.88} & \multicolumn{1}{l|}{0.66} & \multicolumn{1}{l|}{0.75} & \multicolumn{1}{c|}{311} & \multicolumn{1}{c|}{43} \\
\hline
\multicolumn{1}{|c|}{8} & \multicolumn{1}{l|}{0.61}  & \multicolumn{1}{l|}{0.49}  & \multicolumn{1}{l|}{0.55}  & \multicolumn{1}{l|}{0.79}  & \multicolumn{1}{l|}{0.35}  & \multicolumn{1}{l|}{0.48}  & \multicolumn{1}{l|}{1.00} & \multicolumn{1}{l|}{0.88} & \multicolumn{1}{l|}{0.93} & \multicolumn{1}{c|}{315} & \multicolumn{1}{c|}{43} \\
\hline
\end{tabular}
}
\caption{J48, Random Forest, and LibSVM model cross validation results at each iteration.
The size of the training set increases at each iteration as positive predictions made by the model are added to the training set.}
\label{tbl:itrtrain}
\end{table}

\begin{table}[tbh]
\centering
\setlength{\tabcolsep}{1pt}
\scriptsize{
\begin{tabular}{|c|c|c|c|c|c|}
\hline
{Feature name} & \multicolumn{5}{|c|}{Feature rank}  \\
\cline{2-6}
& {SVM} & {ReliefF} & {InfoGain}  & {ChiSquared}  & {Avg.}  \\
\hline
{Sapling-Depth} & {1} & {1} & {1}  & {1}  & {1}  \\
\hline
{Sapling-Number Of Leaves} & {5} & {3} & {3}  & {3}  & {2}  \\
\hline
{Sapling-Balance} & {12} & {2} & {2}  & {2}  & {3}  \\
\hline
{User-Balance} & {7} & {4} & {7}  & {7}  & {4}  \\
\hline
{Sapling-Variety} & {4} & {14} & {4}  & {4}  & {5}  \\
\hline
{Sapling-Number Of Children} & {9} & {15} & {6}  & {5}  & {6}  \\
\hline
{Root-Diversity-50\%} & {10} & {8} & {14}  & {14}  & {8}  \\
\hline
{User-Variety} & {14} & {12} & {10}  & {10}  & {7}  \\
\hline
{User-DisparityNormalized} & {8} & {13} & {13}  & {13}  & {9}  \\
\hline
{Number Of Twigs} & {11} & {20} & {8}  & {9}  & {10}  \\
\hline
{Sapling-Number of Unique Twig Ratio} & {3} & {16} & {16}  & {16}  & {13}  \\
\hline
{Sapling-Number of Unique Term Ratio} & {6} & {10} & {18}  & {18}  & {14}  \\
\hline
{Sapling-Number of Conflicts} & {2} & {22} & {19}  & {19}  & {19}  \\
\hline
\end{tabular}
}
\caption{Feature selection results, with features sorted by their average rank. }
\label{tbl:features}
\end{table}

We trained three different models (J48~\cite{Quinlan:1993:CPM:152181}, Random-Forest~\cite{Breiman01randomforests}, and LibSVM~\cite{Chang01libsvm:a}) on the training set of 200 labeled users, and applied the models to classify unlabeled test data.
We aggregated positive predictions made by all three models, manually labeled them, and iterated the procedure.
Table~\ref{tbl:itrtrain} reports results of cross validation at each iteration.
We reached 100\% precision, 88\% recall and 93\% f-score with LibSVM after eight iterations and stopped at this point.  After eight iterations, our training set had 315 users, of which 43 were experts. Note that only a small fraction of all users can be classified as experts. Self-training enabled us to enrich the training set with positive examples without having to label thousands of users.
Results of 10-fold cross validation of libSVM on  labeled data was 84\% precision, 65\% recall, and 74\% f-score.
Applying the final model to the entire data set identified 66 experts in total.

To see which features are important, we used four feature selection algorithms: SVM Attribute Evaluation~\cite{Guyon:2002:GSC:599613.599671}, Relief  for  Attribute Estimation ~\cite{Kira:1992:PAF:645525.656966}, Information Gain Attribute Evaluation~\cite{Hunt+66}, and Chi Squared Attribute Evaluation~\cite{Greenwood1996}.
SVM Attribute Evaluation method based its decision function on the support vectors of the borderline cases, while others based their decisions on the average cases. This difference leads to different rankings of features. Relief  evaluates the importance of a feature by repeatedly sampling an instance and estimating how well feature values distinguish among instances near each other.
Table~\ref{tbl:features} reports how different features are ranked by these algorithms. All methods identify sapling depth as the most important feature for identifying experts. All methods besides SVM choose the number of leaves in the sapling, and how balanced they are within the sapling, as the next most important features. Generally, sapling-level features are judged to be more important than user-level features by all methods, similar to intuitions of human annotators.

\section{Using Expert Knowledge in Folksonomy Learning}
\label{sec:rap}
Plangprasopchok et al.~\cite{Plangprasopchok11wsdm} proposed a method to learn folksonomies by clustering many saplings created by different users. Their relational affinity propagation (RAP) is a probabilistic method for clustering structured data into a common deeper and bushier tree. RAP merges root nodes of different saplings to extend the breadth of the learned folksonomy, and it merges a child node of one sapling to the root of another to extend its depth.
RAP is based on affinity propagation (AP)~\cite{apscience07}, and it identifies a set of exemplars that best represent all the data. Exemplars emerge as messages are passed between data items, with each item seeking an assignment  to the most similar exemplar. {AP} identifies a set of exemplars, or clusters, which maximize the \emph{net similarity} between  exemplars and data items assigned to them.

\begin{figure}[tb]
\begin{tabular}{@{}c@{ }c@{ }c@{}}
\includegraphics[width=1in]{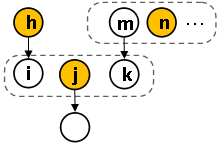}
&
\includegraphics[width=1.4in]{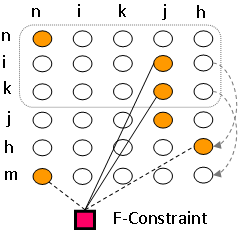}
&
\includegraphics[width=1in]{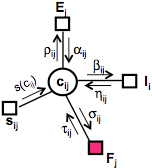} \\
(a) & (b) & (c)\\
\end{tabular}
\caption{Relational affinity propagation ({RAP}): (a) two saplings being merged. Dashed lines surround a group of nodes assigned to the same exemplar (in orange). (b) Binary variable matrix corresponding to the configuration in (a). (c) Factor graph formulation of binary RAP.}
\label{fig:RAP}
\end{figure}

Following binary AP framework of ~\cite{binAP09}, let $c$ be an $N \times N$ matrix, were $N$ is a number of data items. A binary variable $c_{ij} = 1$ if node (data item) $i$ is assigned to node $j$ (i.e., $j$ is an exemplar of $i$); otherwise, $c_{ij} = 0$. AP uses constraints to guide the inference process to ensure cluster consistency. The first constraint,  $I_{i}$, which is imposed on the row $i$,
indicates that a data item can belong to only one exemplar ($\sum_{j}{c_{ij}}=1$). The second constraint, $E_{j}$, which is imposed on the column $j$, indicates that  if an item other than $j$ chooses $j$ as its exemplar, then $j$ must be its own exemplar ($c_{jj} = 1$). {AP} avoids assigning exemplars which violate these constraints.

A similarity function $S(.)$ measures the similarity of a node to its exemplar. If $c_{ij}=1$, then we add $S(c_{ij})$ to the objective function; otherwise, $S(c_{ij})=0$. The self-similarity, $S(c_{jj})$, also called \emph{preference}, is usually set to less than the maximum similarity value in order to avoid creating a configuration with $N$ exemplars. In general, the higher the value of preference for a particular item, the more likely it is to become an exemplar. Setting all preferences to the same value indicates that all items are equally likely to become exemplars.
The global objective function measures the quality of a configuration (i.e., exemplars and items assigned to them):
\begin{eqnarray}
\label{eq:objBinAP}
\textbf{S}(c_{11},\cdots,c_{NN}) & = &  \sum_{i,j}S_{ij}(c_{ij})+\sum_{i}I_{i}(c_{i1},\cdots,c_{iN}) \nonumber \\
& & +\sum_{j}E_{j}(c_{1j},\cdots,c_{1N}).
\end{eqnarray}
A message passing algorithm~\cite{apscience07} is used to find a configuration that maximizes the net similarity  without violating $I$ and $E$ constraints.

\subsection{Relational Affinity Propagation}
In order to cluster structured data into a tree, Plangprasopchok et al.~\cite{Plangprasopchok11wsdm} introduced a new ``single parent'' constraint. The $F$-constraint allows a node to select another as an exemplar only if their parents belong to the same exemplar, thus ensuring that the learned structure forms a tree.
Consider clustering structured data in Fig.~\ref{fig:RAP}(a), where exemplars are in orange, and dashed lines surround nodes assigned to the same exemplar. When child nodes $i$ and $k$ decide whether to merge with node $j$, the $F$-constraint checks whether their parents $h$ and $m$ belong to the same exemplar.
Figure~\ref{fig:RAP}(b) shows the binary variable matrix corresponding the configuration in (a). This configuration is undesirable since it does not correspond to a tree: nodes $i$ and $k$ are assigned to exemplar $j$, but their parents belong to different exemplars.

\remove{
The $F$-constraint can be written as:
\begin{equation}
\label{eq:fconst}
F_{j}(c_{1j},.,c_{Nj}) = \left\{
\begin{array}{l l}
-\infty & \quad \mbox{if $i$ is a child node,} \\
& \quad \mbox{$\exists i: c_{ij} = 1$;} \mbox{$ex(pa(i)) \neq ex(pa(ne(j)))$,}\\
0 & \quad \mbox{otherwise}
\end{array} \right.
\end{equation}
}

\remove{
\begin{figure}[tb]
\begin{tabular}{@{}ccc@{}}
\includegraphics[width=1in]{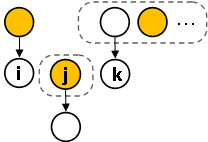}
&
\includegraphics[width=1in]{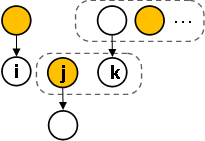}
&
\includegraphics[width=1in]{RAP} \\
(a) & (b) & (c)
\end{tabular}
\caption{Illustration of the $F$-constraint. (a) Original RAP formulation imposes constraints only on child nodes and will lead to a configuration where $i$, $j$, $k$ are in different clusters. (b) Original RAP allows to have  both $i$ and  $k$ are not assigned to $j$.
(c) Factor graph formulation of binary RAP.}
\label{fig:extAP}
\end{figure}
}

In its original formulation, the $F$-constraint was imposed on child nodes only and could result in undesirable configurations.
The $F$-constraint checks whether  $i$ and $k$ can be assigned to $j$, and since they cannot, it forces them into separate clusters. While the configuration is valid, it leads to a shallow folksonomy. We modify the $F$-constraint to prevent such situations.
The modified $F$-constraint is imposed on both child and parent nodes, if the parent node is also an exemplar:
\begin{equation*}
\label{eq:fconst}
F_{j}(c_{1j},\ldots,c_{Nj}) = \left\{
\begin{array}{l l}
-\infty & \quad \mbox{$\exists$ child $i: c_{ij} = 1$ and}  \\
& \quad \mbox{$ex(pa(i)) \neq ex(pa(ne(j)))$}\\
0 & \quad \mbox{otherwise}
\end{array} \right.
\end{equation*}
\noindent where  $ne(.)$ returns a set of nodes that share the exemplar of its argument, $pa(.)$ returns index of the parent of its argument, and $ex(.)$ returns the index of the argument's exemplar.
In the illustration in Fig.~\ref{fig:RAP},  suppose that $k$ is found to be similar enough to $j$ so that they can be merged. To decide whether $i$ too can choose $j$ as an exemplar, the modified $F$-constraint checks whether the parent exemplar of node $i$ is the same as the parent exemplar of any of $j$'s neighbors. If no, $i$ won't be able to pick $j$ as an exemplar.
The objective function in Eq.~\ref{eq:objBinAP} is modified by the addition of the new term $\sum_{j}F_{j}(c_{1j},\cdots,c_{1N})$; we use max-sum method to optimize it.

\subsection{Integrating Expert Knowledge}

RAP provides a framework to integrate experts' knowledge in folksonomy learning. We do this simply by giving the nodes from saplings created by experts higher preference, or self-similarity, values.
This means that these nodes will be more likely to become exemplars, and expert knowledge will guide the folksonomy learning process.

\subsection{Implementing RAP}
Binary RAP may be written as a factor graph shown in Fig.~\ref{fig:RAP}(c).
Following Ref.~\cite{prml06} and  Ref.~\cite{Plangprasopchok11wsdm}, we derived message update formulas for  $\beta$, $\eta$, $\alpha$, $\rho$, $\tau$ and $\sigma$:
\begin{eqnarray}
\label{eq:extbeta}
\scriptstyle \beta_{ij} & = & \scriptstyle s(i,j)+\alpha_{ij}+\tau_{ij},\\
\label{eq:exteta}
\scriptstyle\eta_{ij} & = &   \scriptstyle - \max_{k \neq j} \beta_{ik},\\
\label{eq:extalpha}
\scriptstyle\alpha_{ij} & = & \scriptstyle \left\{
\begin{array}{l}
\scriptstyle\sum_{k \neq j}{\max{[{\rho_{kj}} ,0 }]}\quad {i = j}\\
\scriptstyle\min{[0, \rho_{jj}+\sum_{k \notin {\{i,j\}}} {\max { [ {\rho_{kj} } ,0 } ] } ]} \quad {i \neq j} \\
\end{array} \right.\\
\label{eq:extrho}
\scriptstyle\rho_{ij} & = & \scriptstyle s(i,j)+\eta_{ij}+\tau_{ij},\\
\label{eq:extau}
\scriptstyle\tau_{ij} & = & \scriptstyle \left\{
\begin{array}{l}
\scriptstyle\sum_{k \neq j; k \in S^{\{ ne(j)\}}} {\max{[{\sigma_{kj}} ,0 }]}\quad {i = j}\\
\scriptstyle\min{[0, \rho_{jj}+\sum_{k \notin {\{i,j\}}; k \in S^{\{ne(j)\}}} {\max { [ {\sigma_{kj} } ,0 } ] } ]} \quad {i \neq j} \\
\end{array} \right.\\
\label{eq:sigma}
\scriptstyle\sigma_{ij} & = &  \scriptstyle s(i,j)+\eta_{ij}+\alpha_{ij}.
\end{eqnarray}

In Eqs.~\eqref{eq:extau} and \eqref{eq:sigma}
$\texttt{S}^{\{ne(j)\}}$ represents set of nodes sharing same parent exemplar as neighbors of $j$. Note that we do not need to check all neighbors of $j$, but just one child node among all neighbors, since all nodes in $ne(j)$ must already share parent exemplar. These message update equation will make our model favor the valid configuration, which maximizes the objective function $\textbf{S}(c_{11},\cdots,c_{NN})$.
Since message passing algorithms can be written in max-sum form, they can be easily parallelized on multi-core computers~\cite{map-reduce}. We implemented the message update formulas using map-reduce parallel programming framework~\cite{dean2008mapreduce}, which ran on 30+ node cluster.

\section{Experimental Results}
\label{sec:validation}

We measured the impact of expert knowledge on the folksonomies learned from Flickr data. Our data set consists of 20,759 saplings created by 7,121 users. 
A node can be a collection or a set. The tags of all photos within a set are assigned to the set node and propagated to the collection node. We stemmed all terms (tags, set and collection names) using the Porter stemming algorithm and measured similarity between a pair of nodes $i$ and $j$ by the number of common tags  $t_{ij}$ they have among their top 40 tags: $S(i,j)=min(1., t_{ij}/4)$.
We infer exemplars and clusters by initializing all messages to zero, and update exemplar assignments at each iteration until convergence. We check convergence by monitoring the number of exemplars and the stability of net similarity.

We selected 31 seed terms consistent with Ref.~\cite{Plangprasopchok11wsdm} and generated folksonomies for these seed terms using RAP with and without expert knowledge.
To learn a folksonomy, we first need to select relevant saplings from the data set. We created a snowball sample of relevant saplings as follows. For the seed term that will be the root of the learned folksonomy, first we retrieve all saplings whose root has the same name as the seed term. We then retrieve saplings whose root has the same name as one of the children in the first set of saplings, and so on.
We include expert knowledge in one of two ways: (1) using snowball sample of relevant saplings, including those created by the 66 experts the model identified; (2) in addition to these, use all saplings created by the experts in the snowball sample.

Besides varying the \emph{amount} of expert knowledge used by the learning algorithm,
we can also vary its \emph{weight}. We used the following strategies to vary the emphasis placed on expert knowledge: (1) treat all users uniformly by setting preference values of all nodes to the mean of similarity scores (ordinary RAP); (2) set preference values of expert nodes to twice the mean, while all other preference values are set to the mean.

\begin{figure}[tb]
\begin{center}
\begin{tabular}{@{}c@{}}
\includegraphics[width=0.9\linewidth]{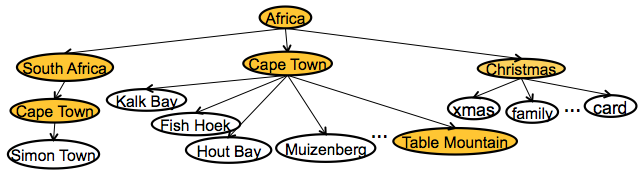}\\
(a)\\
\includegraphics[width=0.9\linewidth]{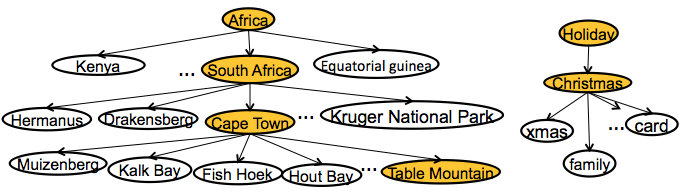}\\
(b)
\end{tabular}
\end{center}
\caption{Folksonomies for `africa' learned (a) without  and (b) with expert knowledge (expert nodes in orange). }
\label{fig:result}
\end{figure}
As an illustration, consider portion of the `Africa' folksonomy, shown in Fig.~\ref{fig:result}(a) learned using saplings such as those in Fig.~\ref{fig:typeSPC}, but without differentiating between expert and novice users. The root has a child `Christmas', because some people spent their Christmas holidays in Africa. Since `Christmas' is linked to many other concepts such as `family', `card', etc,  it introduces irrelevant concepts into `Africa' folksonomy.
Figure~\ref{fig:result}(b) shows portion of the `Africa' folksonomy learned with expert knowledge. Now the $40$ nodes (`xmas', `family', `card', etc.) originally placed under `Africa' $\rightarrow$ `Christmas' were moved to `holiday' $\rightarrow$ `Christmas'. Moreover,  `Table Mountain' and other nodes under `Africa' $\rightarrow$ `Cape Town' were moved under `Africa' $\rightarrow$ `South Africa' $\rightarrow$ `Cape Town'. As we can see from this illustration, adding expert knowledge helps produce a more relevant and detailed folksonomy.

\subsection{Automatic Evaluation}

\begin{table}[tbh]
\centering
\setlength{\tabcolsep}{2pt}
\scriptsize{
\begin{tabular}{|c||c|c|c||c|c|c||c|c|c|c|}
\hline
\emph{seed} & \multicolumn{3}{|c|}{\emph{M1: mean}} & \multicolumn{3}{|c|}{\emph{M2: mean + EXP}} & \multicolumn{4}{|c|}{\emph{M3: 2*mean + EXP}} \\
\cline{2-11}
& \emph{dp}  & \emph{LP} & \emph{TO} & \emph{dp} & \emph{LP} & \emph{TO} & \emph{dp} & \emph{LP} & \emph{TO} & \emph{\%EXP} \\
\hline																					
reptil	&	3	&	0.857	&	0.841	&	3	&	0.857	&	0.8412	&	2	&	1.000	&	0.9199	&	14.28	\\
invertebr	&	3	&	0.197	&	0.599	&	4	&	0.181	&	0.5792	&	4	&	0.183	&	0.6272	&	15.07	 \\
central america	&	3	&	0.134	&	0.586	&	3	&	0.130	&	0.5866	&	3	&	0.130	&	0.5821	&	9.8	 \\
cat	&	3	&	0.024	&	0.587	&	3	&	0.032	&	0.7052	&	3	&	0.472	&	0.8065	&	0	\\
south africa	&	3	&	0.019	&	0.389	&	3	&	0.068	&	0.5022	&	3	&	0.060	&	0.478	&	19.51	 \\
africa	&	3	&	0.396	&	0.610	&	4	&	0.379	&	0.6109	&	4	&	0.457	&	0.671	&	30	\\
craft	&	3	&	0.155	&	0.441	&	5	&	0.263	&	0.486	&	5	&	0.155	&	0.4157	&	1.1	\\
fish	&	3	&	0.079	&	0.335	&	5	&	0.072	&	0.3261	&	3	&	0.174	&	0.4719	&	0	\\
dog	&	4	&	0.014	&	0.496	&	4	&	0.013	&	0.4796	&	4	&	0.020	&	0.6661	&	0	\\
build	&	3	&	0.037	&	0.366	&	5	&	0.003	&	0.3508	&	5	&	0.004	&	0.3714	&	2.77	\\
north america	&	3	&	0.217	&	0.466	&	5	&	0.228	&	0.4116	&	6	&	0.265	&	0.444	&	14.13	 \\
south america	&	3	&	0.095	&	0.416	&	4	&	0.234	&	0.5394	&	4	&	0.292	&	0.5991	&	18.21	 \\
australia	&	3	&	0.171	&	0.541	&	4	&	0.258	&	0.5612	&	4	&	0.179	&	0.5789	&	6.18	 \\
insect	&	5	&	0.027	&	0.349	&	4	&	0.027	&	0.2901	&	4	&	0.032	&	0.3721	&	4.96	\\
flora	&	3	&	0.127	&	0.450	&	3	&	0.127	&	0.4504	&	3	&	0.131	&	0.4523	&	3.52	\\
vertebr	&	4	&	0.034	&	0.390	&	4	&	0.034	&	0.3892	&	3	&	0.273	&	0.5986	&	17.5	\\
urban	&	4	&	0.061	&	0.394	&	4	&	0.061	&	0.3942	&	4	&	0.061	&	0.3946	&	2.64	\\
unit state	&	4	&	0.038	&	0.525	&	4	&	0.038	&	0.5203	&	4	&	0.038	&	0.5236	&	7.93	 \\
bird	&	3	&	0.051	&	0.397	&	5	&	0.052	&	0.3996	&	5	&	0.058	&	0.4497	&	3.97	\\
plant	&	3	&	0.115	&	0.461	&	6	&	0.124	&	0.475	&	3	&	0.351	&	0.584	&	6.25	\\
canada	&	4	&	0.039	&	0.305	&	6	&	0.038	&	0.301	&	4	&	0.075	&	0.4595	&	6.11	\\
unit kingdom	&	3	&	0.219	&	0.583	&	5	&	0.231	&	0.6005	&	6	&	0.216	&	0.5753	&	5.49	 \\
asia	&	4	&	0.052	&	0.449	&	6	&	0.055	&	0.4379	&	5	&	0.056	&	0.4676	&	11.8	\\
sport	&	4	&	0.208	&	0.444	&	6	&	0.226	&	0.4328	&	6	&	0.263	&	0.4575	&	16.46	\\
europ	&	4	&	0.252	&	0.535	&	5	&	0.249	&	0.5333	&	5	&	0.276	&	0.5382	&	10.91	\\
fauna	&	4	&	0.240	&	0.438	&	4	&	0.261	&	0.4448	&	5	&	0.264	&	0.4549	&	11.26	\\
countri	&	4	&	0.075	&	0.530	&	7	&	0.086	&	0.4777	&	7	&	0.118	&	0.515	&	15.11	\\
anim	&	4	&	0.054	&	0.446	&	6	&	0.059	&	0.4328	&	7	&	0.112	&	0.4838	&	9.5	\\
flower	&	5	&	0.053	&	0.391	&	7	&	0.054	&	0.3773	&	7	&	0.053	&	0.3887	&	5.41	\\
world	&	5	&	0.027	&	0.358	&	8	&	0.025	&	0.3137	&	9	&	0.025	&	0.3549	&	17.04	\\
citi	&	5	&	0.005	&	0.500	&	5	&	0.005	&	0.4936	&	5	&	0.007	&	0.502	&	5.67	\\
\hline
\emph{average}	&	3.61	&	0.131	&	0.472	&	\textbf{4.74}	&	0.144	&	0.476	&	4.58	&	 \textbf{0.187}	&	\textbf{0.523}	&	9.44\\
\hline
\end{tabular}
}

\caption{Evaluation of  folksonomies learned for 31 (stemmed) seed terms.}
\label{tbl:re}
\end{table}

Table~\ref{tbl:re} reports results of running RAP in three different settings for 31 seed terms:
(\emph{M1}) relevant saplings collected by the snowball sample with no differentiation between novice and expert users (all preference values set to the \emph{mean}); (\emph{M2}) using relevant saplings plus all other saplings from experts, with no differentiation between users (\emph{mean +EXP}); (\emph{M3}) same saplings as before, but with higher preference values for experts (\emph{2*mean + EXP}). While the learning algorithm generally produces several trees, we evaluate only the most `popular' tree, one that aggregates the greatest number of saplings.
The popular tree learned by M1 contained between 14 and 7925 nodes (2001.26 on average), and that learned by M2 between 16 and 8114 nodes (1947.87 on average), while folksonomies learned by M3 were smaller, between 14 and 5667 nodes (1292.81 on average).

We automatically measure the quality of the learned folksonomies by comparing them to the reference taxonomy from the Open Directory Project (ODP)~\cite{ODP}. We applied two metrics: Lexical Precision (\emph{LP}) and Taxonomic Overlap (\emph{TO})~\cite{dellschaft2006perform}.
LP measures term overlap between the learned and reference taxonomies, independent of their structure,
while TO measures the overlap of ancestors and descendants of a pair of terms from the learned and reference taxonomies without considering their order.
We also measure the \emph{depth} of the taxonomy.
We observe that while RAP leads to few or no structural inconsistencies, integrating expert knowledge into the learning process improves the quality of the learned taxonomies (higher LP and TO scores) and how detailed they are (greater depth), while also removing irrelevant nodes (smaller trees).

Is expert knowledge alone sufficient to produce high quality folksonomies? The last column in Table~\ref{tbl:re} shows the percentage of nodes in the learned folksonomy that can be attributed to experts. On average, this fraction is less than 10\%. We conclude that integrating  knowledge from both expert and novice users leads to more comprehensive folksonomies than using expert knowledge alone.

\subsection{Manual Evaluation}
Automatic method was not comprehensive, since it can only evaluate portions of the learned folksonomies that used the vocabulary of the reference taxonomy. Therefore, we also carried out a manual evaluation using the Coding Analysis Toolkit (CAT)~\cite{CAT}, which provides a Web interface for users to answer customized questions. Each question presented to the user a portion of the learned folksonomy, laid out as a tree, and asked if it was correct. Since the trees were generally very large, we reduced their size as follows. For a pair of folksonomies learned by methods $M1$ and $M3$ for some seed term, we identified leaf nodes with the same name and the same ancestors in the two trees and removed them from both trees. Applying this strategy iteratively eliminated on average 50\% to 70\% of the nodes. If the reduced tree was still large, we segmented it into disjoint subtrees with at most 10 child nodes at any level. We asked five annotators to determine whether each reduced tree (or subtree) was correct (837 questions total). Overall annotators judged 45.30\% of the trees learned by method $M1$ and 68.24\% learned by $M3$ to be correct. Thus, using expert knowledge leads to better folksonomies.

We calculated statistical significance of results of automatic and manual annotation. We find that the difference in TO scores between RAP without and with expert knowledge is significant at 95\% level with t(31)=2.265, p $\le$ 0.05.
Moreover, RAP with expert knowledge improves correctness by 23\% on the manual annotation task. We believe that combining automatic and manual evaluation leads to a convincing evaluation of folksonomy learning.

\subsection{Robustness}
\label{sec:robustness}
\begin{figure}[tbh]
\begin{center}
\begin{tabular}{@{}c@{}}

\includegraphics[width=0.8\linewidth]{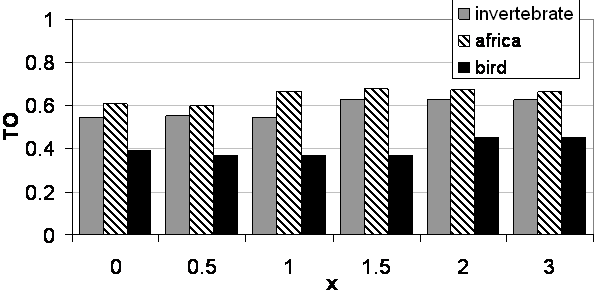}\\
(a)\\
\includegraphics[width=0.8\linewidth]{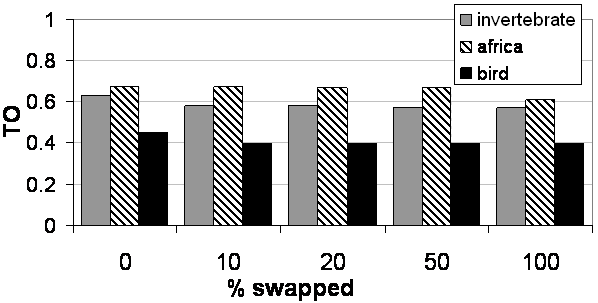}\\
(b)

\end{tabular}
\end{center}
  \caption{Robustness of proposed method, as measured by the taxonomic overlap (TO), with respect to  (a) preference values and (b) percentage of experts misidentified.}\label{fig:robustness}
\end{figure}

Finally, we address robustness of the method with respect to changes in the preference values assigned to expert nodes. We ran our algorithm for six preference values of the form $x^* mean$, where $x\in \{0, 0.5, 1.0, 1.5, 2.0, 3.0\}$.
We report $TO$ scores for three seeds (`invertebrate', `africa', and `bird') in Fig.~\ref{fig:robustness}(left). The quality of the learned folksonomies, as measured by TO, rises with preference values, and saturates around $x=2.0$. 

Another question is how the accuracy of automatic expert identification affects the quality of the learned folksonomies. For this experiment, we randomly selected $n$\% of expert nodes and swapped their preference values with the same number of randomly selected novice nodes. We varied percentage of swapped nodes from 0\% to 100\% and report $TO$ scores for the three learned folksonomies in Fig.~\ref{fig:robustness}(right). As we increased the number of swapped nodes, $TO$ scores dropped by 9\%-12\%. Note that when all expert nodes were swapped for novice nodes, i.e., random novice nodes had their preference values set to $2^*mean$, the $TO$ scores were similar to those that did not differentiate between expert and novice nodes. The difference between 100\% and 0\% swapped is similar to RAP with and without expert knowledge, as expected. We conclude that even moderately high errors (up to 50\%) in expert identification do not significantly degrade the quality of the learned folksonomies.

\section{Related Work}

Expert identification has been addressed by researchers in several different fields.  Existing works analyze the (textual) content of documents people create, the link structure of the interactions between people, or a combination of both methods.
Zhang \emph{et al.}~\cite{zhang2010expert} proposed a probabilistic algorithm to find experts on a given topic by using local information about a person (e.g., publications) and relationships between people.  A similar approach was used by Maybury~\cite{Maybury02} to find experts within organizations from the documents (publications, publicly shared folders) they create and relations between them (project information, citations).
Balog \emph{et al.}~\cite{balog2006formal} used generative language models to identify experts among authors of documents, while Deng \emph{et al.}~\cite{deng2009formal} explored topic-based model for finding experts in academic fields. Davitz \emph{et al.}~\cite{Davitz}  used network analysis tools to identify experts based on the documents or email messages they create within their organizations. Content quality analysis in social media has been investigated from many research.  Agichtein \emph{et al.}~\cite{Agichtein08findinghigh-quality} investigated methods to measuring quality of contents by content, user relationship features.  Hu \emph{et al.}~\cite{Hu_abstractmeasuring} proposed quality accessing model using the interaction data between articles and their contributors.
Our approach is similar in spirit, in that we look at the contents of data people create to identify experts, although we have not yet included relations between people into analysis. Unlike these earlier methods, we use the structure of annotations to measure their expertise on a topic. While Korner \emph{et al.}~\cite{koerner2010stop} proposed a method to differentiate users in social tagging systems, they classify users as categorizers and describers based on their tag usage, and show that there is more semantic agreement between describers. They do not attempt to learn taxonomies nor  differentiate the quality of annotations.

With the advent of crowdsourcing services, labeling large datasets has become easier. However, due to variations in annotators' abilities, significant post-processing is required. To address this problem, Welinder \emph{et al.}~\cite{WelinderPeronaCVPR2010} proposed a labeling strategy based on the estimation of most likely value of current labels and annotator's abilities. Sheng \emph{et al.}~\cite{Sheng_getanother} studied repeated--labeling strategies to improve label quality. Our work is different in the sense that on the Social Web users freely choose content to label, as well as labels themselves (tags, directories), that reflect their own interest in content.
Our work is also related to broader  efforts to ``crowdsource'' knowledge production, embodied, for example, by  ``citizen science'' projects and ``wisdom of crowds'' approaches~\cite{Steyvers09}. Researchers have studied methods that aggregate data of varying quality~\cite{Hemmer10,Jun10icdm}. However, the amount and variation of data in these studies was limited. Our approach can automatically identify the quality of data and aggregate it from thousands of users.

\section{Conclusion}
In this paper, we propose a framework to automatically identify experts based on the linguistic and structural features of the annotations they create, and use experts' annotations to guide the folksonomy learning process. We show that using experts' knowledge can produce more accurate and detailed folksonomies. We also show that proposed method is robust to errors in expert identification. Our work generalizes beyond Flickr to other structured data sources (eBay categories, Delicious bundles, Bibsonomy relations, file systems).

In  future work, we would like to extend automatic expert identification procedure using Bayesian approach~\cite{Steyvers09}. Experts are able to be modeled in continuous variable rather than 1 or 0 binary variable. By identifying experts in more detail, we could control the degree to which experts knowledge is used.
We would also like to extend RAP to apply to other structure learning problems, such as alignment of biological data. Finally, we would like incorporate more efficient inference algorithm and compare the aproach to other statistical relational learning approaches.

\section*{Acknowledgment}
We would like to thank Jong-hyop Kim for implementing CAT evaluation and the annotators for evaluating the learned folksonomies. This work is based on research supported by the National Science Foundation under awards IIS-0812677 and CMMI-0753124.

\IEEEtriggeratref{9}


%
\bibliographystyle{IEEEtran}

\end{document}